# White Paper on High Temperature Superconducting Bi-2212 Magnets for Energy Frontier Circular Colliders

March 15th 2022

T. Shen[1]

[1] Lawrence Berkeley National Laboratory, Berkeley, CA 94720

A.V. Zlobin[2]

[2] Fermi National Accelerator laboratory, Batavia, IL 60510

D. Larbalestier[3]

[3] The Applied Superconductivity Center at the National High Magnetic Field Laboratory at Florida State University, Tallahassee, FL 32310

Abstract:

As the only high-temperature superconducting (HTS) material available as an isotropic, twisted, multifilamentary round wire, Bi-2212 is very promising for expanding the high-field superconducting accelerator magnet toolbox beyond the round-wire, isotropic Nb-Ti and $Nb_3Sn$ conductors used by HEP so far. In this paper, we describe the important roles that Bi-2212 might play for future high energy circular colliders including both high energy proton and muon colliders. We describe its present technology status (conductor development, magnet design concepts, prototype magnet status) and then provide a ten-year plan for >15 T accelerator magnets and >25 T solenoid magnets within an integrated strategy to engage industry and the entire US and international scientific enterprise interested in HTS magnet applications.

# Table of Contents

# 1. Introduction and Opportunity

High field superconducting magnets are used in particle colliders [1-3], fusion energy reactors [4], magnetic resonance imaging (MRI) scanners, ion beam cancer therapy devices [5], as well as thousands of nuclear magnetic resonance (NMR) and general laboratory magnets. So far, virtually all superconducting magnets have been made from two Nb-based low temperature superconductors - Nb-Ti with superconducting transition temperature $T_c$ of 9.2 K and Nb$_3$Sn with $T_c$ of 18.3 K. Our high-field magnet tool box could be greatly expanded by high-$T_c$ cuprates and iron-based superconducting materials with upper critical magnetic field ($\mu_0 H_{c2}$) exceeding 50 T at 4 K, much greater than that of Nb-Ti (~14 T at 1.8 K) and Nb$_3$Sn (~26-27 T at 1.8 K). For particle colliders, the usefulness of achieving high magnetic fields is readily seen from the simple description of the proton beam energy of proton colliders $E \propto B \times R$, where $B$ is the field in the dipole magnets (T), and $R$ is the bending radius (km) and the particle energy $E$ in TeV. Coincidently, magnetic fields of >25 T generated by superconducting magnets are also desired by condensed materials research[4], fusion energy[5], and NMR[6], whereas >30 T superconducting solenoids are important for providing the final cooling needed by a multi-TeV Muon collider to decrease the transverse beam emittance by orders of magnitude for final acceleration and collision [6, 7].

For HEP circular colliders, the 8.33 T Nb-Ti accelerator dipole magnets of the large hadron collider (LHC) at CERN enabled the discovery of the Higgs Boson and the ongoing search for physics beyond the standard model of high energy physics, whereas 12 T class Nb$_3$Sn quadrupole magnets are key to the high-luminosity upgrade of the LHC that aims to increase the luminosity by a factor of 5–10 [8, 9]. Superconducting accelerator magnets for circular colliders generate powerful magnetic fields for HEP experiments but often suffer from field errors due to persistent magnetization currents in the superconducting filaments and/or eddy currents within a non-twisted superconducting strand and within a superconducting cable. In Nb-Ti magnets, field errors are minimized by embedding the superconducting filaments within a normal metal matrix, reducing the filament diameter to <10 $\mu$m, and twisting the strand. Eddy currents are controlled by twisting the superconducting filaments within the strand and twisting, transposing and strand coating when the interstrand resistance within a cable must be controlled. The Rutherford cable is the essential path to a high current cable for all the Nb-Ti and Nb$_3$Sn accelerator magnets made so far. It is a flat rectangular cable that is simply manufactured from flattening round strands, which is readily manufactured and capable of delivering accelerator field quality.

The superconducting magnet field limit has a strong influence on the design of circular colliders like FCC-hh or CPPC or muon colliders and the reach of physics. The development of high-temperature superconducting (HTS) magnets for frontier particle physics colliders was endorsed by the 2014 P5 report and its 2015 accelerator R&D subpanel report and by the 2020 update of European strategy for particle physics [10]. Since the 2014 P5, significant progress has been made by developing accelerator magnets from two HTS, REBCO coated conductors and Bi-2212 round wires. Unlike Nb-Ti, Nb$_3$Sn, and Bi-2212, REBCO coated conductors are produced only in tape forms with a large shape aspect ratio, for which Rutherford cable fabrication is not applicable. Moreover REBCO tapes are single filaments with no filament twisting, and with a strong anisotropy of $T_c$, $H_{c2}$, and critical current density $J_c$ in magnetic fields. The broad apes have large persistent current magnetizations, especially in perpendicular magnetic fields, which is typically two orders of magnitude larger than desired.

However, REBCO has a magnetic field reach which is unmatched by any other practical superconductor and CERN has pursued an aligned block dipole design based on a flat ROEBEL cable assembled from meandered-shape punched REBCO tapes [11, 12]. The aligned block dipole design largely keeps the magnetic field parallel to the tape plane (at least at the straight section) and thus keeps field errors under control. A drawback is that the ROEBEL cable requires cutting away ~40% of the coated conductor, significantly raising cost and potentially increasing the mechanical vulnerability of the cable against thermal and powering cycles. In the US, the US magnet development program (MDP) has been developing a canted cosine theta dipole based and traditional cosine theta dipoles both based on CORC®, a round cable made with helically wound REBCO tapes around a Cu core. For CORC® magnets, the magnetic field thus sees all orientations of the tape, lowering the attainable $J_c$ and also making magnetic field quality control a challenge. Additional challenges include improving the technical maturity of CORC® cable to meet electrical (inter-strand resistance control to minimize eddy currents), mechanical (damages associated with bending and handling reacted conductors and the weak interlayer bonding inherent in a multiple layer thin film) and developing quality control and assurance methods [13-16].

By contrast Bi-2212 is the only high-$T_c$ cuprate material available as an isotropic, twisted, multifilamentary round wire [17] that can use long-length Rutherford cable technology, opening up the same magnet geometries developed for the Nb-Ti and Nb$_3$Sn magnets. Bi-2212 is thus applicable to block-type, cosine theta [3], canted cosine theta (CCT) [18], and a recently developed stress management cosine theta [19] construction schemes. Being a multifilamentary round wire with small magnetization, it also has a strong potential for building >23 T solenoid magnets needed by ultra high field (1.2-1.5 GHz, 28-36 T) NMR magnets and muon colliders. The wire engineering critical current density $J_E$ has been steadily improved beyond that needed for constructing practical superconducting magnets. And magnet design concepts have been proposed and tested in short subscale magnets with good field quality. The ongoing research will be further described in following sections.

The goal of this document is to describe the technology status and define appropriate steps to develop Bi-2212 into a practical very high field (>15 T for accelerator magnets, and >25 T for solenoids) magnet technology, from superconductor development, to design and development leading to the construction of model coils and magnets fulfilling accelerator-quality needs for muon or hadron colliders. The first and most relevant target is to develop high-field accelerator dipole magnets capable of generating 16-20 T in a bore of 40-50 mm suitable for colliders. The second target is to develop large-bore, high-field >15 T quadrupole magnets for interaction regions. A third target is to develop 25 T or greater commercial solenoid magnets in collaboration with the US magnet industry and to develop the essential elements of the technology needed to build >30 T solenoids for muon colliders. Given the potential helium shortage, a fourth relevant target is to develop Bi-2212 magnets, both solenoids and accelerator magnets, for other special uses at higher working temperatures (10-25 K). The paper first explains the general challenges to building superconducting accelerator magnets, then moves to describe specific challenges of building Bi-2212 magnets, then describes ongoing research, and finally outlines a ten-year plan with opportunities for collaboration and industry partnerships.

## 2. Superconducting Accelerator Magnets: Requirements and Challenges

The widely used superconducting accelerator magnet technology is based on a two-layer, cosine-theta coil wound from Rutherford cable in a roman-arch structure [2, 3]. The first key test of the suitability of superconductors for accelerator magnets is whether the winding current density $J_w$ can be made sufficiently high (>400 A/mm$^2$) for efficient field generation. A primary requirement is a wire engineering current density $J_E$ of >600 A/mm$^2$ (the total current carried by a superconducting strand divided by its total cross-section area). A cable with low porosity and thin insulation is also essential.

Second, it is important to have mechanical structures to manage the tremendous Lorentz forces that compress cables azimuthally towards the mid-plane of the winding and radially outwards. The 4 T Nb-Ti Tevatron magnet has an outward force of 11.8 kN per meter magnet length. Both forces go up with $B^2$. Nb-Ti is ductile. However, Nb$_3$Sn, Bi-2212, and REBCO are brittle [20]; the dependence of their superconducting properties on mechanical loads needs to be well understood and ingenious engineering solutions need to be found.

Third, it is also important to minimize training, and have reliable circuit protection during quenches in place. A magnet can be destroyed by its stored energy, which goes up with $B^2$, if the energy is deposited within a small fraction of the superconducting winding. In addition, during quench, a large electrical voltage to ground is generated and therefore a robust insulation is required. Nb-Ti and Nb$_3$Sn magnets suffer from quench training, which refers to the increase of peak current observed in a magnet when it undergoes a series of tests where the current is ramped up until the magnet quenches. HTS conductors offer a potentially training-free superconducting magnet, but also present challenges that need new solutions.

Accelerator magnets operate in cycling mode with various field ramps. The Large Hadron Collider (LHC) main-ring dipole magnet is cycled from injection (0.54 T) to a high field (8.3 T) in a period of 20 mins, maintained at high field for 2-10 hours for beam collision, and then finally back to low field for another cycle, where it may remain for 20-40 minutes during which particles are injected. Magnets need to provide great accuracy of the magnetic field seen by particles over the full dynamic range from field injection through collision. The error fields need to be of the order of $10^{-4}$ of the central bending field for stable beam behavior, and thus of the order of a few 0.1 mT. To provide field accuracy, it is necessary to control the geometry of the conductor and winding (change of any of dimensions should be constrained to the order of 0.025 mm), and controlling field distortions due to eddy currents that flow between strands during and following a field ramp. In addition, superconductors come with a persistent current that shields superconductor filaments from the external field. Such persistent currents generate spurious fields, especially $b_3$ (sextupole) at low field. The implication is that the magnetization of the superconductor needs to be minimized. Magnetization scales as , where $D_{eff}$ is the effective diameter of superconductor filaments, and therefore $D_{eff}$ needs to be small (6-7 μm for LHC main dipole wires). The RRP® Nb$_3$Sn strand selected for the high luminosity LHC (HL-LHC) quadrupole QXF magnets has a $D_{eff}$ of 55 μm.

## 2.1. Magnet Technology Challenge

In this section, we discuss magnet technology challenges that are specific to Bi-2212. Preliminary data show that Bi-2212 Rutherford cables can show 5% critical current degradation under ~120 MPa transverse pressure [22]. A recent 20 T dipole magnet concept design within the US MDP has shown that it is possible to control the transverse stress on the Bi-2212 cable to <110 MPa [21]. To increase the safety margin, two stress management accelerator magnet concepts have been proposed, theCanted Cosine Theta (CCT) at the LBNL and the Stress Management Cosine Theta (SMCT) at the Fermilab. They have been previously described in detail here [22-25] and [28]. The CCT is particularly friendly to manufacture, which has allowed much recent prototyping work, as will be described in next sections.

Like $Nb_3Sn$, Bi-2212 magnets are fabricated using a wind and react process that requires no further bending or extra handling after the conductor is reacted and sensitive to overstrain. A key advance for Bi-2212 since the 2014 P5 is the introduction of the overpressure processing heat treatment, which uses a high-pressure mixed Argon and $O_2$ (50 bar typically with 98% Ar and 2% $O_2$) to densify the superconducting filaments and obtain a high wire $J_E$. For single wires a $TiO_2$ coat on the wire and a braided mullite insulation works well but this is not yet working as well for Rutherford cables because an application of a robust $TiO_2$ has not yet been achieved. The result is some Bi-2212 leakage through the encasing metal (Ag or Ag alloy) at high temperatures which can react with surrounding $SiO_2$-containing insulation [26, 27]. Leakage degrades wire $J_E$ because it depletes a portion of filaments and interrupts current flow. Methods to evade these restrictions are under active development.

Superconducting magnets have a large magnetic stored energy. When superconductors lose superconductivity locally due to thermal or mechanical or electromagnetic disturbances, it is very important to distribute the stored energy as uniformly as possible in the coil winding or to extract as much energy as possible into room temperature apparatus, typically into a dump resistor. Nb-Ti and $Nb_3Sn$ magnets are prone to quench due to a low $T_c$ and enthalpy margin. On the other hand, it is much easier to spread out the stored energy for Nb-Ti and $Nb_3Sn$ magnets due to fast normal zone propagation velocity (~10s m/s in Nb-Ti and $Nb_3Sn$ versus ~1-10s cm/s for HTS conductors) [26]. Slow growth of normal zones in HTS magnets makes reliable quench detection critical to reliable quench protection. Although quench detection by voltage measurements is always possible in principle, late quench detection is common due to slow normal zone propagation which can eat much of the safety margin for quench protection. This challenge is certainly not unique to Bi-2212 magnets and the challenges are much higher with REBCO magnets.

To fabricate accelerator magnets of >15 T and solenoids of > 25 T, HTS coils are combined with Nb-Ti and $Nb_3Sn$ to minimize overall costs. Typical designs use HTS inserts nested concentrically inside Nb-Ti and/or $Nb_3Sn$ outsert magnets, setting up issues of magnet assembly and integration for both accelerator and solenoid magnets given their very different stabilities and quench velocities. The US MDP design indicates that a 20 T dipole magnet will require a six-layer superconducting coils, with the inner two layers supplied by Bi-2212 and the outer four layers by Nb$_3$Sn. The HTS magnets, in standalone configuration, will have to generate a dipole field of ~10 T whereas Nb$_3$Sn magnets, in standalone configuration, will have to generate a dipole field of ~15 T. This magnet integration will require careful modeling, design, and experimental validation.

The main ring dipole magnets for circular colliders range from 5-15 m long. Scaling the magnet fabrication from ~1 m long demonstration magnets to 5-15 m presents another challenge. The assembly between coils and between coils and structures is much easier to handle for 1 m long magnets. The coil reaction will require building heat treatment facilities that are sufficient to handle such long magnets. The magnet stored energy goes up linearly with the magnet length and thus quench protection may become an issue. Such a challenge is not unique to Bi-2212 but the complex reaction and fabrication detailed above certainly adds into the challenges.

## 2.2. Industrialization Challenge

A present challenge of Bi-2212 wires is the high cost. Bi-2212 wires use silver as a metal matrix. Over the last 5 years, the silver cost has fluctuated from $490/Kg to $980/Kg, large but still significantly smaller than the Bi-2212 powder cost of ~$3000/Kg. For wires containing about 20% Bi-2212, the raw material cost is then about $1100/Kg using the highest price for Ag. Present selling price of the final wire is of order $15,000/Kg, giving a final to raw material cost ratio, normally called the P ratio, of about 13.5, well above the mature production values of Nb-Ti and Nb$_3$Sn wires [28]. The 'production scaling factor' *P* decreases systematically with billet mass. With further industrial development, a potential *P* of 3 is possible with manufacturing processing towards maturity and large scale [28]. That places the final cost of the Bi-2212 wires in the range of $3300/Kg (2022 currency), at the level of today's Nb$_3$Sn RRP® wires that are used for the high-luminosity LHC upgrade. This is in fact not far from the present selling price of Bi-2223 tape made by Sumitomo Electric who make it by essentially the same route as Bi-2212 is made and sell it on a much larger scale. Price can decrease with driving down precursor powder cost.

A risk that is not unique to Bi-2212 presents a clear and present danger. The long term survival of the US superconductor industry might be at risk over the next decade. For one reason, the US superconductor industry is facing fierce competition from low cost oversea superconductor manufacturers, in particular those in China, which leverages the growth of local medical market (MRI) and participation in the ITER program and has developed a competitive Nb-Ti industry. Nb-Ti superconductor is a commodity and advanced superconductors like Nb$_3$Sn and Bi-2212 are where the US still holds a strong edge. The competitive margin of the U.S. industry will shrink, if without support or the US market needs to develop advanced superconductors. What is at stake is more than the survival of the US superconductor industry. Superconducting magnets are important for various scientific missions across the US DOE offices whereas the strength of the US MRI manufacturers will significantly decrease without a strong local supply of superconductors. What is particularly relevant is that Bi-2212 wires are manufactured by the Bruker OST LLC at Carteret, NJ, which also manufactures the RRP® Nb$_3$Sn wires for the high-luminosity LHC upgrade.

## 3. Ongoing Activities

### 3.1. Conductor

Commercial Bi-2212 wires are available with a diameter that ranges from 0.8 mm to 1.5 mm and a piece length of 2 km (0.8 mm Ø) sufficient for (5-15 m long) final HEP facility accelerator magnets. In the US, Bi-2212 wires are fabricated by Bruker OST LLC, Carteret, NJ, and the precursor powder is supplied by the Engi-Mat LLC, Lexington, KY. In 2017, high engineering current density of 1000 A/mm$^2$ at 4.2 K and 27 T has been demonstrated [29, 30]. Filament twisting was shown at industrial scale and in model magnets. It is worth mentioning that even such high $J_c$ values are only at about 1% of the depairing current densities that set the fundamental limits of critical current density in a superconductor [31]. For Nb-Ti and REBCO, the realized $J_c$ is about 20% of the corresponding depairing current densities. Potential reasons identified and being worked on now are extensive filament distortions (sausaging) during manufacture and imperfect connectivity across grain boundaries after heat treatment [32].

Much of the conductor characterization and materials science work is supported by the university GARD grants from the US DOE-OHEP annual competitions [29, 31-34]. Key messages from recent work are that (1) the industrially supplied conductor and powder are consistent, as judged by application of a standard 50 bar overpressure heat treatment (OPHT). It is no longer necessary to individually optimize each wire to get a predictable high $J_E$. Point (2) is that the OPHT, whether done at short sample scale or coil scale (presently up to 44 cm length and 14 cm diameter but soon (Q2-2022) with a new furnace 1 m long and 25 cm in diameter) is routine because consistent as judged by more than 15 coil runs over the last 2 years. Point (3) is that we are gaining increasing confidence from extensive long sample comparisons that measure the distribution of $I_c$ values along the length of 1 mm wires now being delivered in 1 km lengths fairly routinely that they are also rather uniform from a superconducting point of view too. Extensive study of the distribution of filament diameters in the as-delivered state shows significant sausaging which only gets worse during the high temperature portions of the heat treatment. Point (4) is that when we compare the already high $J_c$ values obtained by dividing $I_c$ values by the Bi-2212 cross-sections, we see that only about 1% of the depairing current density is obtained, some 20 times less than occurs in optimized Nb-Ti and REBCO coated conductors [31]. Present collaborations with B-OST are aimed at exploring the manufacture of much more uniform filaments whose reaction degradation can be better controlled. We finally note (5) that the uniaxial tension degradation of Bi-2212 wires is very progressive, irreversible degradation starting at strains between 0.4 and 0.45% and then degrading slowly at strains up to 50% larger, values quiet compatible with internal strengthening methods being used for high field solenoids. Compressive degradation property measurements are being set up in an ASC-FSU/LBNL/U. of Twente collaboration. In short there has been huge progress in Bi-2212 technology since P5 2014 centered on work jointly performed at the ASC-FSU-NHMFL and LBNL with strong industrial collaborations with B-OST, Engi-Mat, and most recently with Cryomagnetics in a joint effort with ASC-FSU-NHMFL to build a quasi-commercial 25 T hybrid solenoid magnet.

## 3.2. High Field Dipoles

Since the 2014 P5, the Berkeley lab (LBNL) has collaborated with the NHMFL, under the framework of the U.S. MDP since 2016, and wire industry, additionally supported by the US DOE SBIR-STTR program, to fabricate flat racetrack coils and CCT coils using Bi-2212 Rutherford cables. Applying the overpressure processing and leveraging wire $J_E$ improvement, they quadrupled the quench current of flat racetrack coils fabricated to 8.6 kA (with a 3.5 T field) [30, 35]. Valuable experience has been gained with wire fabrication, cable design, insulation, coil winding, heat treatment, and test. The best wire $J_E$ achieved in racetrack coils is 1000 A/mm$^2$ at 4.2 K and 3.6 T [30]. A gap is identified between the coil (both racetrack and CCT) performance and short strand performance [22, 25, 30, 35].

Four CCT coils were fabricated and tested. Their field quality was modeled and measured. A CCT dipole magnet was assembled and tested in 2021. The CCT magnet BIN5c1, which used a 9-strand Rutherford cable, achieved 3.6 kA with a dipole field of 1.64 T in a bore of 30.8 mm. BIN5c1 is the world's first Bi-2212 dipole magnet with a clear bore. Importantly, none of the racetrack coils and CCT coils, either with or without epoxy impregnation, showed quench training or degradation due to electromagnetic or thermal cycles. The BIN5c1 magnet also exhibits fast ramping capability (1 kA/s and 0.46 T/s) and low field hysteresis (<1 mT at 1 T, hall probe measurement). The Bi-2212 CCT magnets thus have gone through the full magnet design, fabrication cycle from wire, to cable, to magnet fabrication, and test and ready to be scaled up both in length (from ~40 cm for BIN5 magnets to 1 m long Bi-CCT1 and Bi-CCT2 CCT dipole magnets) and magnetic field (from 2.5 T BIN5 to ~5 T for Bi-CCT1, and 6.5 T Bi-CCT2) with wider cables of larger current carrying capability.

Both CCT and SMCT face interesting magnet fabrication related issues. First is Coil winding and structural components fabrication. Both CCT and SMCT coils use winding mandrels of unique geometries that are fabricated using 4-axis CNC machining and advanced manufacturing. It is interesting to adopt additive manufacturing mandrels with novel metals and Nickel based alloys with tolerances that are compatible with generating sufficient field quality for accelerator magnets.

Since 2018, Fermilab has been pursuing a SMCT dipole design for $Nb_3Sn$ and Bi-2212 in collaboration with LBNL and NHMFL-FSU under the framework of the U.S. US-MDP [36]. Whereas the LBNL group pursues coils based on a Canted Cosine Theta (CCT) concept, Fermilab is focusing on traditional cos-theta coils with stress management elements. The Fermilab's Bi2212 insert coil is based on a 2-layer coil concept with small aperture and outer diameter (OD) to fit into 60 mm aperture $Nb_3Sn$ dipole coils [37]. It uses Rutherford cable with Bi2212 strands and rectangular cross-section. The cable is wound inside a support structure produced using fast-developing 3D Advanced Manufacturing technology. After winding, the coil is reacted to form Bi2212 stoichiometry at 50 bar pressure at NHMFL, and shipped back for epoxy impregnation, instrumentation, assembly and testing at Fermilab. The calculated conductor limit of the individually powered Bi2212 coil in the mirror structure is 4.1 T at 8 kA, and in the 4-layer hybrid mirror of 8.5 T at 7 kA.

The next Bi2212 coils will use advanced Bi2212 wires produced after 2017. These coils will be assembled and tested in the 4-layer hybrid mirror or in the 4-layer hybrid dipole configurations. It is expected that the maximum field of the individually powered Bi2212 coil in the mirror will increase to 5.8 T at 12 kA. In the 4-layer hybrid mirror with the Bi2212 coil connected in series with the Nb3Sn coil, the maximum field will increase to 11 T at 10 kA.

### 3.3. Solenoid magnets

A central component of the ASC-FSU-NHMFL effort has been to develop Bi-2212 as a high field solenoid technology, first for general purpose 25 T laboratory solenoids that are also of interest for muon accelerators, and then for service towards making 1.5 GHz NMR spectrometers that require about 36 T field with order $10^{-7}$ homogeneity in 1 cm$^3$ or better. To make this possible has required development of internal stress management techniques that are presently proprietary but sufficiently interesting to be attracting strong commercial licensing interest. A STTR collaboration to build a 25 T solenoids with dual use as both a muon and a laboratory solenoid should come to fruition in late 2022 or early 2023.

## 3.4. Facilities and Technical Support

Two facilities are crucial for the development of Bi-2212 magnets. The overpressure processing heat treatment furnace is needed for the reaction. Since 2015, a DELTECH furnace at the NHMFL has been used to heat treat about two dozens of solenoid and accelerator magnet type coils (racetrack and CCT). All of these coils' heat treatments were successful, confirming that such a process can be feasible and reliable. Much engineering experience has been gained. DETECH uses a hot-wall design that limits its hot zone to 45 cm long and 130 mm in diameter. A new overpressure processing heat treatment facility, RENEGADE, is being commissioned at the NHMFL by the same team. Using a cold wall design (the pressure vessel works at a temperature no larger than 250 °C), RENEGADE is designed to have a hot zone, within which temperature varies less than 1 °C, of 250 mm in diameter and 1.2 m in length. This facility will enable the reaction of 1 m long prototype CCT and SMCT coils.

The second facility that is important for developing $Nb_3Sn$-HTS hybrid magnets is a test facility with capability to provide two independent power circuits with sufficient circuit protection and strategies to manage the magnetic coupling and quench protections that may arise in a hybrid magnet configuration. Such facilities clearly need investment and ideally are possessed by multiple labs where coil fabrication takes place. A synergy is with the U.S. DOE fusion office, which has funded with the U.S. DOE OHEP to construct a 15 T test facility dipole (with aperture of 94 mm x 144 mm) [38] at Fermilab for testing HTS fusion cables. This facility is expected to have capability to provide a variable temperature environment for assessing HTS cables' capabilities at temperatures other than 1.8 K or 4.2 K.

# 4. Ten Years Plan

We now come to the following recommendations as paths to developing Bi-2212 for accelerator applications :

(1) All Bi-2212 magnets depend on a stable supply of good conductors with reliable properties. Thus it is essential to continuously understand the properties of each 10 kg (presently costing about $150 K but with hopes of reducing by factors of 5-20 in next 10 years as demand and experience increase). As for Nb-Ti and Nb$_3$Sn, an essential component of the background R&D is to understand the conductor, especially present limitations on properties so that ongoing collaborations in the university-industry-lab nexus can drive further improvements. This work will be ongoing throughout the whole magnet development program so as to do everything possible to bring Bi-2212 to the same level of development as today's Nb$_3$Sn conductors.

(2) Use CCT and SMCT Bi-2212 magnets as R&D vehicles to drive conductor development and accelerator magnet design, technology, and test towards maturity. The goal is to build multiple short (up to 1 m long) dipole model magnets with a bore of 40-50 mm in standalone configuration with incremental dipole field from 5 T to 10 T. Several critical technology elements are to be optimized and developed: a) Master technology knowhow of the reaction of ~1 m long coils and demonstrate reliability. b) Optimize Bi-2212 Rutherford cable fabrication including its packing pactor and insulation, especially to minimize the contribution of insulation materials to the thermodynamic ceramic leakage.

(3) In hybrid magnet configuration, fully explore the hybrid magnet technology and its reproducibility using three test beds, including Fermilab's mirror test configuration, the 120 mm bore, 11 T US MDP Nb$_3$Sn dipole magnet to test hybrid magnets (independently powered) in 12-15 T and use the 15 T fusion HTS cable test facility to test insert coils (independently powered) to generate dipole fields of >16 T. Two independent power circuits provide the flexibility for testing magnets fabricated from cables with various current carrying capability. Gain experience with magnet operation and quench protection for the new kind hybrid magnets.

(4) With technology maturing, build multiple, 1 m long >16 T, series-connected (one power circuit) Nb$_3$Sn/Bi-2212 hybrid accelerator dipole magnets with a projected bore diameter of 40-50 mm. Gain fabrication and operation experience. A critical assessment is whether accelerator magnet performance and field quality can be realized and reproduced.

(5) Perform a critical assessment in the feasibility of scaling up accelerator magnets (technologies and infrastructure, etc.) to 5-15 m long. A task is to assess feasibility and develop plans for infrastructure development for coil reaction.

(6) Engage the possibility of developing interaction region Bi-2212 quadrupole magnets.

(7) Use 25 T commercial solenoid magnets as another vehicle to drive conductor development and solenoid magnet technology.

(8) For both accelerator magnets and solenoid magnets, explore special uses at higher working temperatures (10-25 K).

(9) Engage the US superconductor and magnet industry to develop production capabilities for large scale raw materials, wire and cable fabrication, and to reduce cost and supply chain risk.

## 5. Opportunities for Collaboration

### 5.1. Other DOE Offices and NSF

As mentioned, the HTS fusion cable test facility being funded by both the US DOE fusion office and the US DOE OHEP at Fermilab offers a dipole field of 15 T with an aperture of 94 mm x 144 mm that can house future Bi-2212 HEP dipole coils and enables exploration and demonstration of dipole magnets at 15-20 T. In addition, the fusion energy community is moving towards building compact fusion reactors for which HTS magnets are a key technology. While current effort is focusing on the toroidal field (TF) coils that form the plasma bottle using REBCO coated conductors. TF coils are working in steady states and suit REBCO coated conductors for which AC losses are high. For the central solenoids which are pulsed, seeking a higher field beyond those established by the ITER central solenoids will require using HTS conductors; for this application, multifilamentary, low AC loss Bi-2212 should be considered. The efforts on high field solenoids supported by NSF are a clear opportunity of collaboration and synergy.

### 5.2. International Collaboration

Domestic collaboration that occurs within the US MDP since 2014 P5 has been a key driver of Bi-2212 magnet technology. Collaboration with Europe with the formation of the High Field Magnet (HFM) program will add access to tools and expertise in Europe. An example is the cable transverse pressure testing at the University of Twente and the FRESCA test facility at CERN.

## 5.3. Industry

Industry involvement in the process of superconducting wire and precursor powder fabrication has been vital and a strength of the U.S. superconductor and superconducting magnet programs. We strongly encourage and support collaboration between national labs, universities, and wire and magnet industry under U.S. DOE SBIR-STTR programs and accelerator stewardship and accelerator R&D and production (ARDAP) to develop innovations and address industry issues such as cost and supply chain risk reduction.

## 6. Summary

Success in developing HTS high field dipole magnet technology to further maturity at 1 m length and 20 T field levels will bring real choice for the design of next proton and muon colliders. We have summarized the present status and plans for applying Bi-2212 to high field magnet technology for both accelerator magnets and for broader solenoid magnet applications in the commercial sector.

ACKNOWLEDGEMENT

We thank colleagues at our respective institutions for their contributions to the R&D work discussed in this document, especially Laura Garcia Fajardo, Ray Hafalia Jr., Diego Arbelaez, Lucas Brouwer, Shlomo Caspi, Stephen Gourlay, Paolo Ferracin, Maxim Marchevsky, Cory Myers, Ian Pong, Soren Prestemon, Gianluca Sabbi, Reed Teyber, Xiaorong Wang at the LBNL, Emanuela Barzi and Igor Novitski at the FNAL, Daniel Davis, Ulf Trociewitz, Eric Hellstrom, Jianyi Jiang, Fumitake Kametani, Youngjae Kim, Ernesto Bosque, the Applied Superconductivity Center PhD students, Najib Cheggour, and Lance Cooley at the NHMFL. We are also grateful to industrial partners who have made critical contributions to wire fabrication and magnet development, including Bruker OST LLC, Engi-Mat LLC, and Cryomagnetics Inc.

DISCLAIMER

The views, as well as errors, in this paper belong solely to the authors, and do not reflect the official view of the U.S. Magnet Development Program.